\documentclass[12pt,preprint]{aastex}
\usepackage{epsfig}
\usepackage{natbib}
\begin{document}
\title{SDSSJ115517.35+634622.0: A Newly Discovered Gravitationally Lensed Quasar}
\author{
Bart~Pindor,\altaffilmark{1}
Daniel~J.~Eisenstein,\altaffilmark{2}
Naohisa Inada,\altaffilmark{3}
Michael~D.Gregg,\altaffilmark{4,5}
Robert H. Becker,\altaffilmark{4,5}
Jon~Brinkmann,\altaffilmark{6}
Scott Burles,\altaffilmark{7}
Joshua~A.~Frieman,\altaffilmark{8,9,10} 
David E. Johnston,\altaffilmark{8,9}
Gordon T. Richards,\altaffilmark{1}
Donald~P.~Schneider,\altaffilmark{11}
Ryan Scranton,\altaffilmark{12}
Maki Sekiguchi,\altaffilmark{3}
Edwin~L.~Turner\altaffilmark{1}
and Donald~G.~York,\altaffilmark{8}
}

\altaffiltext{1}{Princeton University Observatory, Peyton Hall, Princeton, NJ 08544.}
\altaffiltext{2}{Steward Observatory, University of Arizona, 933 North Cherry Avenue, Tucson, AZ 85721.}
\altaffiltext{3}{Institute for Cosmic Ray Research, University of Tokyo, 5-1-5 Kashiwa, Kashiwa City, Chiba 277-8582, Japan.}
\altaffiltext{4}{Physics Department, University of California, Davis, CA 95616.}
\altaffiltext{5}{IGPP-LLNL, L-413, 7000 East Avenue, Livermore, CA 94550.}
\altaffiltext{6}{Apache Point Observatory, 2001 Apache Point Road, P.O. Box 59, Sunspot, NM 88349-0059} 
\altaffiltext{7}{Physics Department, Massachusetts Institute of Technology, 77 Massachusetts Avenue, Cambridge, MA 02139.}
\altaffiltext{8}{Department of Astronomy and Astrophysics, The University of Chicago, 5640 South Ellis Avenue, Chicago, IL 60637.}
\altaffiltext{9}{Center for Cosmological Physics, The University of Chicago, 5640 South Ellis Avenue, Chicago, IL 60637.}
\altaffiltext{10}{Fermi National Accelerator Laboratory, P.O. Box 500, Batavia, IL 60510.}
\altaffiltext{11}{Department of Astronomy and Astrophysics, The Pennsylvania State University, University Park, PA 16802}  
\altaffiltext{12}{Department of Physics \& Astronomy, University of Pittsburgh, 3941 O'Hara St., Pittsburgh, PA 15260}

\bibliographystyle{aj}

\begin{abstract}

We report the discovery of SDSSJ115517.35+634622.0, a previously unknown gravitationally lensed quasar. The lens system exhibits two images of a $z = 2.89$ quasar, with an image separation of $1{\farcs}832 \pm 0.007$ . Near-IR imaging of the system reveals the presence of the lensing galaxy between the two quasar images. Based on absorption features seen in the Sloan Digital Sky Survey (SDSS) spectrum, we determine a lens galaxy redshift of $z = 0.1756$. The lens is rather unusual in that one of the quasar images is only $0{\farcs}22\pm0{\farcs}07$ ($\sim 0.1 R_{\rm eff}$) from the center of the lens galaxy and photometric modeling indicates that this image is significantly brighter than predicted by a SIS model. This system was discovered in the course of an ongoing search for strongly lensed quasars in the dataset from the SDSS. 

\end{abstract}

\section{Introduction}

Under certain conditions, the gravitational potential of a foreground galaxy can produce multiple images of a background quasar. These rare events are of considerable interest due their utility in a number of astrophysical investigations, including modeling the mass distribution of the lens galaxy \citep{ 1995ApJ...445..559K}, measuring the Hubble constant through time delay measurements \citep{1964MNRAS.128..307R,1997ApJ...482...75K,1999ApJ...527..513K}, and constraining cosmological models through the use of lensing statistics \citep{1990ApJ...365L..43T, 1990MNRAS.246P..24F,2002PhRvL..89o1301C}. Many of these applications would benefit from a sample of gravitational lenses defined by a clearly understood selection procedure, so as to allow for the correction of selection effects. 

The Sloan Digital Sky Survey \citep[henceforth SDSS]{2000AJ....120.1579Y} dataset provides an excellent opportunity to discover gravitationally lensed quasars. The $\sim 10^4$ deg$^2$ of five-band photometry and the spectroscopic sample of $\sim 10^5$ quasars should allow for the compilation of a lensed quasar sample which is both large and statistically well-defined in comparison to existing optical samples. For further discussion on this subject, the reader is referred to Pindor et al. (\nocite{2003AJ....125.2325P}2003, henceforth P03). In this work, we report the discovery of a previously unknown gravitationally lensed quasar, found during our ongoing search for such objects in the SDSS dataset. 

\section{Observations}

\subsection{Selection as a Lens Candidate} 

Five-band photometry of SDSS~J115517.35+634622.0 (henceforth SDSS~J1155+6346) was obtained in the course of normal SDSS imaging\footnote{The SDSS photometric designation is (run/rerun/camCol/field/id) 2078/22/1/84/169} on 2001 January 26. The object was classified as a galaxy by the automated star-galaxy separator and is bright enough ($r_{\mathrm{PETRO}} = 16.8$) that it was targeted for spectroscopy by SDSS galaxy target selection \citep{2002AJ....124.1810S}. SDSS spectroscopy\footnote{The SDSS spectroscopic designation is (mjd/plate/fibre) 52316/598/87} on 2002 February 11 instead revealed the object to be a quasar at a redshift of $z = 2.89$. Having been identified as a quasar, the object was initially selected as a lens candidate by a method similar to that used to select the first two SDSS-discovered gravitational lenses (\nocite{Inada1,Inada2}Inada et al. 2003a, Inada et al. 2003b) as well as for having apparent \ion{Mg}{2} absorption at a redshift of $z = 0.72$. It finally received follow-up spectroscopy when observed as part of a sample which passed the selection criteria described in P03. 

Briefly, both of the aforementioned selection algorithms work by searching among spectroscopically confirmed quasars for objects that are extended\footnote{We use the word ``extended'' to mean ``not consistent with being a point source'' and the word ``resolved'' to mean ``displaying two distinct photometric peaks''} and that can be decomposed into two components of similar colours. The algorithm of Inada et al. uses photometric model likelihoods measured by the SDSS photometric pipeline \citep[henceforth PHOTO]{2001adass..10..269L} to identify extended objects and decomposes objects using flux moments. P03 uses photometric models consisting of one and two point spread functions (henceforth PSF) to identify extended objects and to decompose objects. Our two PSF photometric model of the SDSS atlas image predicted a component separation of $1\farcs87$, a flux ratio of A/B = 5/4, and a position angle of $-87^{\circ}$.

	The SDSS is a photometric and spectroscopic survey across 10,000 square degrees of the northern Galactic cap using the 2.5m SDSS telescope at Apache Point Observatory. SDSS imaging is carried out with a wide-field camera \citep{1998AJ....116.3040G} which makes nearly simultaneous observations of objects in five passbands: {$u$ $g$ $r$ $i$ $z$}. Together, the passbands cover the optical wavelengths from the atmospheric cut-off in the blue to the minimum detectable energy for the silicon CCDs in the red \citep{1996AJ....111.1748F}. Photometric calibration of the imaging survey is separately carried out by an automated 0.5m telescope which monitors a set of standard stars \citep{2002AJ....123.2121S} while photometric data is being acquired\citep{2001AJ....122.2129H}. The SDSS imaging camera also incorporates astrometric CCDs which provide astrometry of detected objects with an accuracy typically better than 0$\farcs1$ \citep{2003AJ....125.1559P}. SDSS spectroscopy is carried out on the same telescope by two fiber-fed double spectrographs which produce spectra with a resolution ($\lambda / \Delta \lambda$) of $\sim$ 2000 covering the wavelength range 3800--9200 \AA. Together, the spectrographs have 640 fibers which are assigned based on previous SDSS imaging through an efficient tiling algorithm \citep{2003AJ....125.2276B}. For more comprehensive documentation of the survey, readers are encouraged to consult \citet{2002AJ....123..485S} and \citet{Abazajian03}.

\subsection{Spectroscopic Observations}

	We observed the object on 2002 May 4, using the 6.5m Multiple Mirror Telescope (MMT). Spectra were obtained using the blue channel of the MMT spectrograph with the 300/mm grating and $1^{\prime\prime}$ slit. The slit was aligned along the separation axis of the pair, as predicted from SDSS imaging by a photometric model consisting of two point sources (see P03), so that the two components were observed simultaneously. The spectrum was obtained with a single 300s integration. 
  
	Although the seeing at the time of observation was slightly sub-arcsecond, the two objects were only marginally resolved in the focal plane due their small separation ($\sim 1{\farcs}8$). Consequently, we implemented a deblending procedure to minimize cross-contamination between the extracted 1D spectra. We binned the 2D spectrum into 20 bins along the dispersion axis and integrated the flux in these bins to produce a series of spatial cross-sections. We then simultaneously fit a profile consisting of two Gaussians to each of these cross-sections, subject to the constraint that the separation of the two Gaussians be the same in each bin. Finally, we defined two apertures by assigning a fraction of the flux from each pixel to the first (second) aperture corresponding to the fractional contribution of the first (second) Gaussian to the model profile at that pixel. Hence, the model profiles can be thought of as relative weights which determine how much of the flux in a given pixel should be assigned to either of the two apertures. This deblending procedure is conceptually based upon the deblending algorithms used by PHOTO. Having been thus deblended, the spectra were then extracted using standard IRAF routines. Figure 1 shows the spectra of the two components. Note that the feature at $\sim$ 7600 {\AA} is atmospheric. We cross-correlated the spectra and estimated the velocity difference between them to be $100 \pm 300$ km~s$^{-1}$, consistent with zero.  

We further examined the spectra for evidence of emission from a lens galaxy. The SDSS spectroscopic pipeline identifies in the SDSS spectrum a cross-correlation match to a series of galactic absorption features at a redshift of $z = 0.1756$. In fact, this match is reported with a confidence level (99.73\%) only marginally lower than the match to the quasar redshift (99.75\%). This redshift identification is slightly suspect due to the fact that the purported calcium $H$ and $K$ lines appear in the Lyman-alpha forest. However, as shown in Figure 2, a number of the absorption features are clearly visible in both the SDSS and MMT spectra. Hence, we report a redshift measurement of $z = 0.1756$ for the possible lens galaxy. 

We also examined the absorption doublet seen at $\sim$ 4810 {\AA} in both components. This doublet was identified in the SDSS spectrum as being \ion{Mg}{2} in absorption at a redshift of $z = 0.72$. Its location relative to the broad \ion{N}{5} emission line might suggest that it is the \ion{N}{5} doublet in absorption in the vicinity of the quasar. In Figure 3 we show the best-fit locations of the \ion{Mg}{2} and \ion{N}{5} doublet systems overlayed on the SDSS spectrum, which, although spatially unresolved, has better spectral resolution and signal-to-noise than our MMT spectrum. We conclude that we are in fact observing \ion{Mg}{2} absorption at a much lower redshift. It is possible that this absorber is associated with a mass concentration which contributes to the lensing properties of the system, but it is impossible to test this hypothesis with the existing data. 

We consider the spectra alone to be inconclusive with respect to whether this system was a lensed or binary quasar. The lensing hypothesis is supported by the fact that the spectra are similar in overall shape and have similarly peaked emission lines. Spectroscopic identification of a possible lens galaxy clearly also supports the lensing hypothesis. On the other hand, the equivalent widths of the emission lines in the fainter B component are about twice the widths in the brighter A component. This seems inconsistent with the spectra being two images of the same quasar. However, in the next section, we present further evidence that reconciles the observed spectral differences with the lensing hypothesis.     

\subsection{Near-IR Imaging}   

	Having determined that SDSS~J1155+6346 did indeed have two separate quasar images at the same redshift and with similar SED, we next obtained a near-IR image. On 2002 August 8, we imaged the object in the $K^{\prime}$-band using the Near Infrared Camera (NIRC, Matthews \& Soifer 1994 \nocite{1994ExA.....3...77M}) on the Keck I telescope. The observations consisted of a five-point dither pattern, integrating for 10s at each pointing. The data were flattened, sky subtracted, shifted, and stacked using the DIMSUM package in IRAF.  

	Figure 4 shows the $K^{\prime}$-band of SDSS~J1155+6346. It is evident from visually inspecting this image that component A is extended. It is our interpretation of this image that SDSS~J1155+6346 is indeed a gravitational lens, and that the lensing galaxy is almost co-incident with the quasar image we have labelled A. To support this interpretation, we first constructed a model of the system consisting of only two point sources. We used an analytic PSF consisting of two Moffat functions, as recommended by \citet{1996PASP..108..699R}. The half-width of our PSF model was determined by fitting to a nearby star, labelled P. Figure 5a shows the residuals of the best-fit two point source model. The peak residuals are 32\% of the peak intensity. The highly negative (white) residuals near the center of component A are produced as the model attempts to account for the extended profile by over-estimating the central flux. Since the images are background-dominated, the measurement errors can be inferred from the observed sky variance, and our two point-source model has a reduced chi-square of 442. We subsequently constructed a model of the system as a combination of two point sources plus an extended component represented by a deVaucouleurs profile of the form: 

\begin{eqnarray}
I(x,y) = I_0 \mathrm{ exp }(-7.67((x^2 + (y/q)^2)^{1/2}/R_{\rm eff})^{1/4})
\end{eqnarray}
convolved with the PSF, where $q$ is the axis ratio ($b/a$). In general, our model should also allow the position angle of the galaxy on the sky to vary, but we found that the major and minor axes happen to be well-aligned with the pixel grid. Our best-fit model has a reduced chi-square of 6 (with 11014 degrees of freedom) and the peak residuals are 7\% of the peak intensity. We interpret the obvious improvement observed for our two point sources plus deVaucouleurs model as definitively demonstrating the presence of a lens galaxy. Figure 4b shows the residuals left after subtracting our model from the data. The structure of the residuals near component B indicate that the inaccuracy of our PSF model probably dominates the uncertainties in the model. For this reason, estimation of the errors in the model parameters is not straightforward. We estimated the random errors through bootstrap re-sampling, but this does not adequately represent the systematic errors. Hence, we considered two further photometric models; one consisting of two analytic PSFs (as before) plus an exponential disc, and one consisting of two empirical PSFs, directly using the image of P as a PSF model, plus a deVaucouleurs profile (as before). The best-fit exponential disc model has a reduced chi-square of 21, and it differs from our initial model mainly in that it predicts roughly equal flux in the two quasar images and substantially less total flux in the lens galaxy. Figure 5c shows the residuals left after subtracting this model from the data. The substantially worse chi-square value indicates that the lens is an early-type galaxy, as is in fact indicated by visual inspection of the image. The best-fit empirical PSF model actually has a lower reduced chi-square value, 4.5, than our initial model. However, using the image of P directly requires by-hand masking of a fainter nearby source, as well as of the local gradient produced by the extended profile of the lens galaxy. For this reason, we retain our initial model as the favoured photometric model of the system. Figure 5d shows the residuals left after subtracting the empirical PSF model from the data. The main difference between the results of the empirical PSF model and our initial model is that the distance between the center of the lens galaxy and quasar image A [henceforth $\Delta \theta_{AG}$] is about twice as great in the former as in the latter. Hence, the main systematic uncertainty appears to be in the relative position of the lens galaxy, and the extent of this uncertainty is indeed underestimated by our bootstrap error estimation. Our favoured model predicts $\Delta \theta_{AG} = 0{\farcs}15$, but a more robust estimate can be obtained from average of $\Delta \theta_{AG}$ as predicted by the analytic and empirical PSF models, whereby $\Delta \theta_{AG} = 0{\farcs}22\pm0{\farcs}07$ (systematic). The fairly large reduced chi-square of our best-fit model indicates that further information (i.e. more reliable errors) could be obtained from the image if a more satisfactory PSF model could be found. Table 1 summarizes the best-fit parameters for all three models. The near coincidence of the lens galaxy and quasar A can lead to some confusion so, for clarity, let us explicitly state: when refering to, for instance in Figure 1, component A, we mean the unresolved combination of quasar A and the lens galaxy, which is brighter than the resolved image of quasar B. However, the results of our photometric modeling indicate that quasar image A itself is fainter than quasar image B.   

	In order to further verify the results of our photometric model, we constructed a difference spectrum by multiplying the spectrum of component A by the flux ratio predicted by our model and then subtracting the result from component B. Figure 6 shows this difference spectrum. The difference in \ion{C}{4} line appears to be consistent with zero, but significant residuals remain in the vicinity of Lyman-alpha. However, given that our sightline to quasar image A obviously passes quite close the center of the lens galaxy, it is not unreasonable to expect some difference in the quasar spectra due to absorption and/or micro-lensing. Further, if our identification of the lens galaxy redshift, $z = 0.1756$, is correct, then the residuals are quite possibly due to the 4000 {\AA} break which we would expect to observe in the vicinity of $\sim$ 4700 {\AA}. We interpret the difference spectrum as broadly confirming the results of our photometric modelling, but as not having sufficient signal-to-noise to contribute to estimation of the lens model parameters. 
  
	We also matched SDSS~J1155+6346 with the 2MASS Second Incremental Data Release Point Source Catalog and found it to have a $K$-band magnitude of 14.6$\pm 0.1$. We can use the 2MASS measurement to calibrate the total magnitude of the system, so that our photometric model of the NIRC image predicts $K$ magnitudes\footnote{We neglect the conversion between $K$ and $K$' as it is a  negligible correction relative to our photometric accuracy.} of 14.9$\pm 0.1$, 16.6$\pm 0.1$ and 17.4$\pm 0.1$ for the galaxy, quasar B, and quasar A, respectively.

	A measurement of the lens galaxy magnitude allows us to estimate its redshift, using the Faber-Jackson (\nocite{1976ApJ...204..668F}1976) relation for lensing as presented by \citet {Rusin02}. We re-write Rusin et al.'s equation (A1) in the form

\begin{eqnarray}
m_{obs} - M_{*0} = DM + 2.5 \gamma_{E+K}z_d - 1.25\gamma_{FJ}\mathrm{log}\Delta\theta_{red}
\end{eqnarray}
where $DM$ is the distance modulus, $\gamma_{E+K}$ is a parameter subsuming evolution and spectral K-corrections, $\gamma_{FJ}$ is the slope of the Faber-Jackson relation in the chosen band, $z_d$ is the lens galaxy (deflector) redshift, and $\Delta\theta_{red}$ is the reduced image separation, $\Delta\theta_{red} \equiv (\Delta\theta / \Delta\theta_{*})(D_s/D_{ds})$, where $\Delta\theta_{*}$ is the image separation produced by a singular isothermal sphere with a velocity dispersion of 225 km~s$^{-1}$, and $D_s$ and $D_{ds}$ are the angular diameter distances between the observer and source and deflector and source, respectively. The $K$-band values of $M_{*0} (=-24.4 + 5 \mathrm{log}h)$, $\gamma_{E+K} (=-0.31)$, and $\gamma_{FJ} (=3.40)$ are all given by Rusin et al., based upon HST photometry and spectrophotometric modelling of 29 early-type lens galaxies. We can then numerically solve Equation (2) to get $z_{galaxy} = 0.17 \pm 0.03$. Hence, our photometric redshift estimate is in excellent agreement with the tentative spectroscopic redshift presented in \S 2.2. If we accept the galaxy redshift to be $\sim$ 0.17, then an SIS model predicts a velocity dispersion of $\sim$ 190 km~s$^{-1}$ in order to reproduce the observed image separation. However, an SIS model also predicts a B/A flux ratio of $\sim$ 10:1 (for our favoured model, $\sim$ 5:1 for the galaxy position in the empirical PSF model), rather than the observed $\sim$ 2:1. Possible explanations for this discrepancy include some combination of a large internal or external shear, microlensing, or a significant softening of the inner slope of the mass profile. The discrepancy could also be explained if the distance between the galaxy and quasar image A is significantly greater than our models predict.     

\section{Conclusions}

In the course of an ongoing search for strongly lensed quasars in the dataset produced by the SDSS, we have discovered an object which is almost certainly a previously unknown gravitationally lensed quasar. High angular resolution spectroscopy confirms two quasar images with an angular separation of $\sim 1{\farcs}8$ having identical redshifts of $z = 2.89$. We also identify the presence of absorption features corresponding to a lens galaxy at a redshift of $z = 0.1756$. High resolution near-IR imaging clearly indicates the presence of a luminous, early-type galaxy between the quasar images. Our photometric model of the system implies an angular separation of $1{\farcs}832 \pm 0.007$ between the quasar images. We used 2MASS photometry to measure the lens galaxy magnitude as being $K=14.9$. This magnitude allows for a photometric estimate of the lens galaxy redshift as being $z = 0.17 \pm 0.03$, based on the Faber-Jackson relation for lensing. One of the quasar images is only $0{\farcs}22\pm0{\farcs}07$ ($\sim 0.1 R_{\rm eff}$) from the center of the lens galaxy. Hence, the optical depth to quasar micro-lensing \citep{2001glrp.conf..185W} for this image may be by quite high. Similarily, higher resolution multi-band imaging of the system could provide an interesting measurement of the differential reddening \citep{1999ApJ...523..617F} produced by the central region of the prominent lens galaxy. 

B.P.\ is supported by NASA grant NAG5-9274. D.J.E.\ is supported by National 
Science Foundation grant AST-0098577 and by an Alfred P. Sloan Research Fellowship. Part of the work reported here was done at the Institute of Geophysics and Planetary Physics, under the auspices of the U.S. Department of Energy by Lawrence Livermore National Laboratory under contract No.~W-7405-Eng-48. G.T.R.\ is supported by HST grant HST-GO-09472.01-A. 

A portion of the results reported here made use of the Multiple Mirror Telescope Observatory, a facility operated jointly by the University of Arizona and the Smithsonian Institution. 

Additional observations reported here were obtained at the W. M. Keck Observatory, which is operated as a scientific partnership among the California Institute of Technology, the University of California, and the National Aeronautics and Space Administration. The observatory was made possible by the generous financial support of the W. M. Keck Foundation.

This publication makes use of data products from the Two Micron All Sky Survey, which is a joint project of the University of Massachusetts and the Infrared Processing and Analysis Center/California Institute of Technology, funded by the National Aeronautics and Space Administration and the National Science Foundation.

Funding for the creation and distribution of the SDSS Archive has been provided by the Alfred P. Sloan Foundation, the Participating Institutions, the National Aeronautics and Space Administration, the National Science Foundation, the U.S. Department of Energy, the Japanese Monbukagakusho, and the Max Planck Society. The SDSS Web site is http://www.sdss.org/. 

The SDSS is managed by the Astrophysical Research Consortium (ARC) for the Participating Institutions. The Participating Institutions are The University of Chicago, Fermilab, the Institute for Advanced Study, the Japan Participation Group, The Johns Hopkins University, Los Alamos National Laboratory, the Max-Planck-Institute for Astronomy (MPIA), the Max-Planck-Institute for Astrophysics (MPA), New Mexico State University, University of Pittsburgh, Princeton University, the United States Naval Observatory, and the University of Washington.

\begin{figure}
	\centering\epsfig{figure=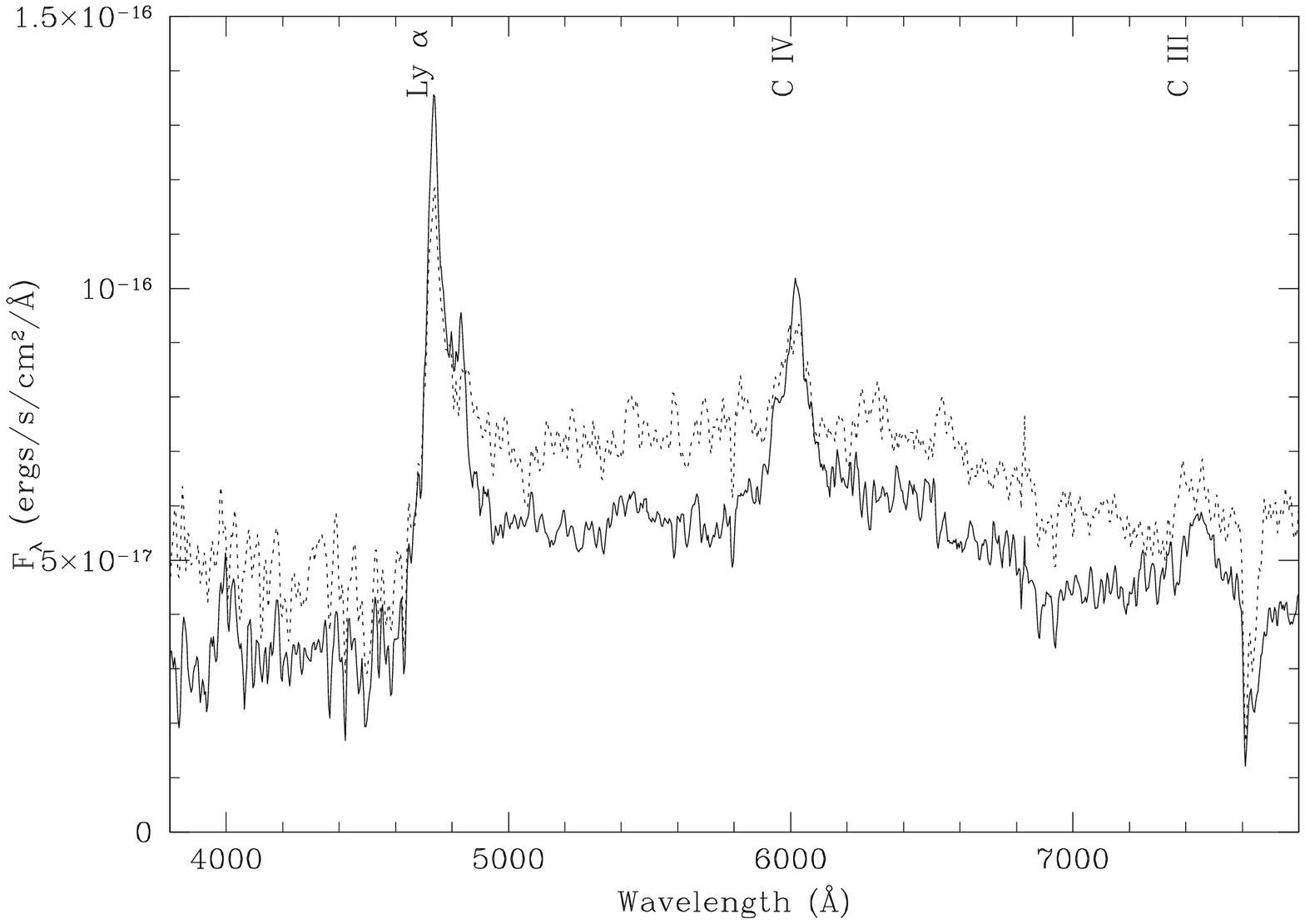,width=\textwidth}
	\caption{Superimposed MMT spectra of component A (dashed line) and component B (solid line). Vertical labels indicate the positions of quasar emission lines at a redshift of $z = 2.89$. The feature at $\sim$ 7600 {\AA} is atmospheric.}
\end{figure}

\begin{figure}
	\centering\epsfig{figure=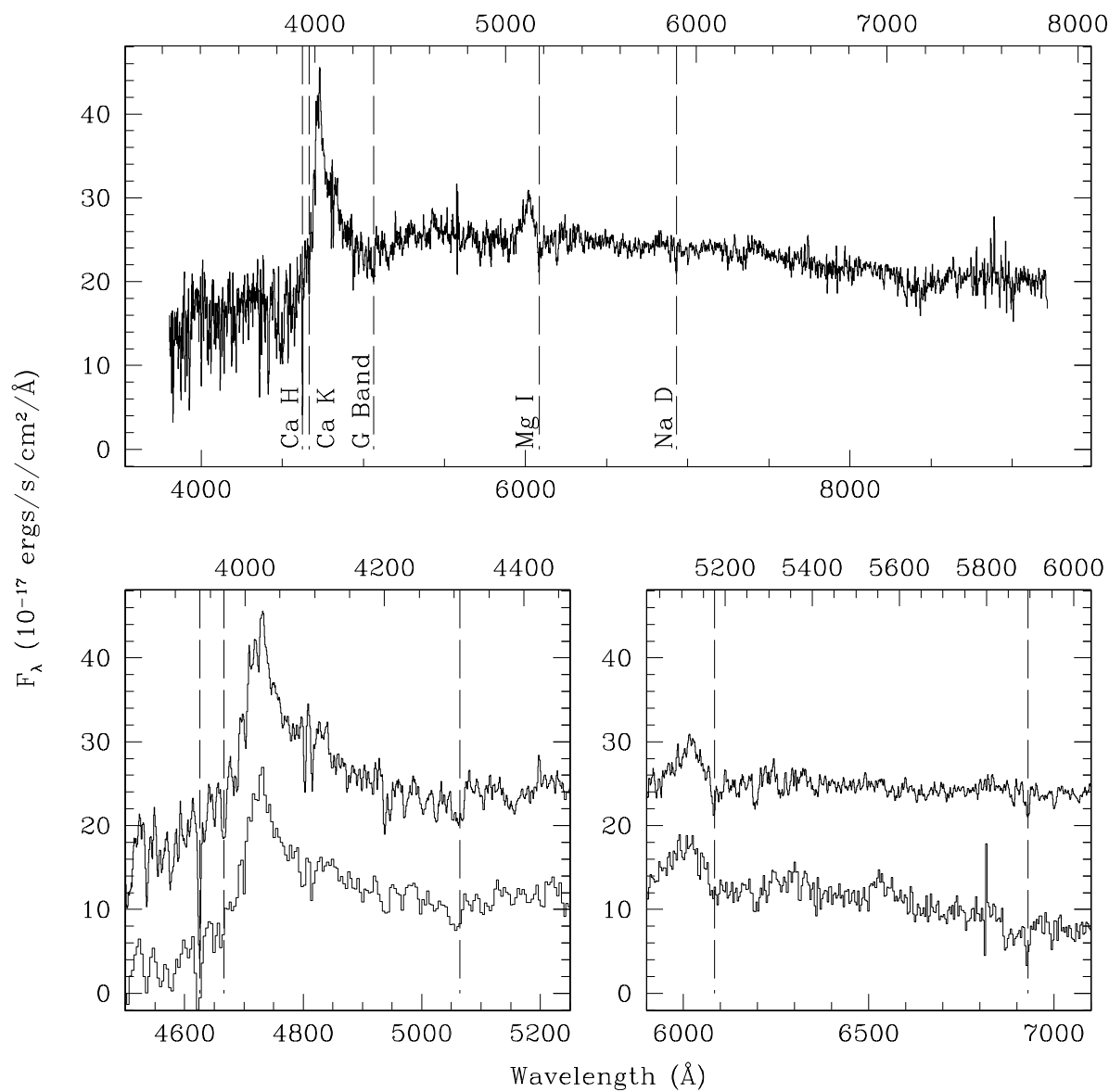,width=\textwidth}
	\caption{Upper Panel: SDSS spectrum of SDSS~J1155+6346 with a template of prominent galactic absorption lines at redshift 0.1756 superimposed. The indicated transitions are (rest wavelengths, in \AA): Ca $K$ (3934) $H$ (3969) G band (4308) \ion{Mg}{1} ($\sim$ 5175) Na $D$ ($\sim$ 5895). Lower Panels: Close-up on regions with possible absorption features associated with the lens galaxy. The upper spectrum is the unresolved SDSS spectrum. The lower spectrum is the MMT spectrum of component A. The spectra have been scaled to have the same counts and an arbitrary offset has been introduced to separate them. In all panels, the lower wavelength scale denotes the observed wavelengths and the upper scale denotes rest wavelengths corresponding to a redshift of 0.1756. }
\end{figure}

\begin{figure}
	\centering\epsfig{figure=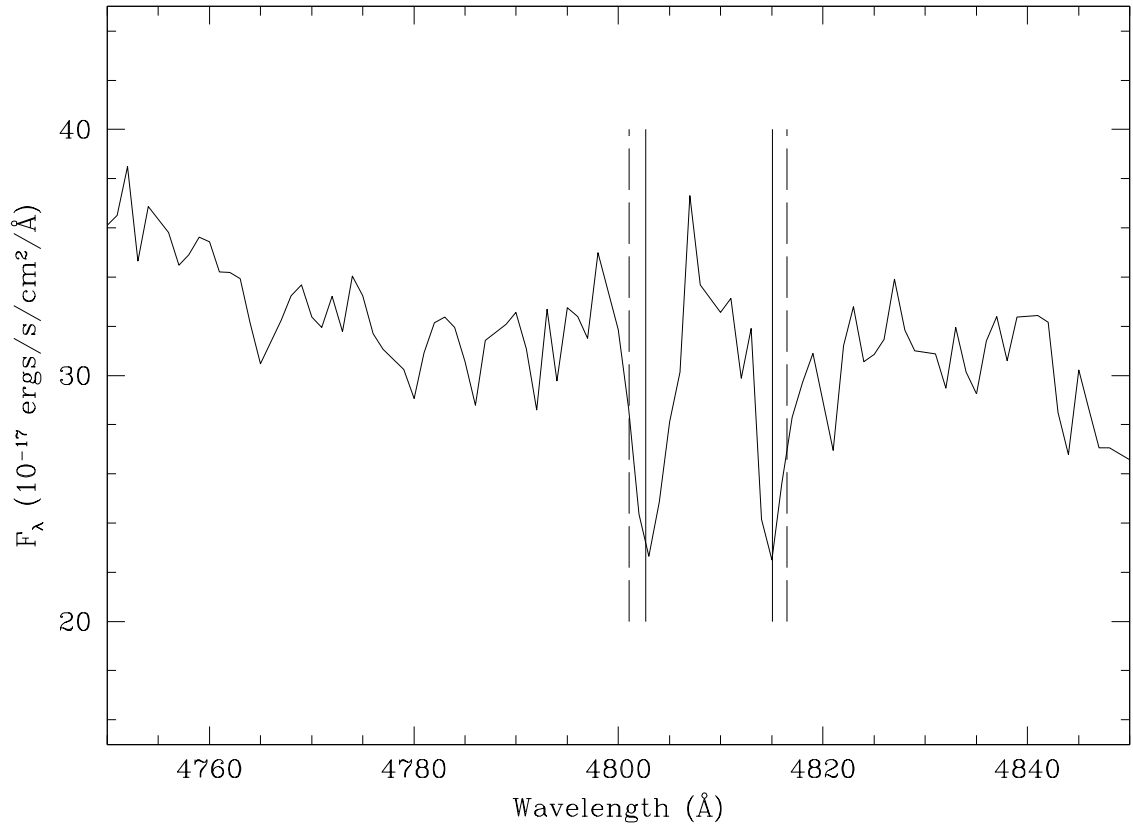,width=\textwidth}
	\caption{SDSS spectrum of SDSS~J1155+6346 in the vicinity of 4800 {\AA}. The solid vertical lines indicate the position of the \ion{Mg}{2} doublet at a redshift of $z = 0.718$. The dashed lines indicate the position of the \ion{N}{5} doublet at a redshift of $z = 2.8755$}

\end{figure}

\begin{figure}
	\center\epsfig{figure=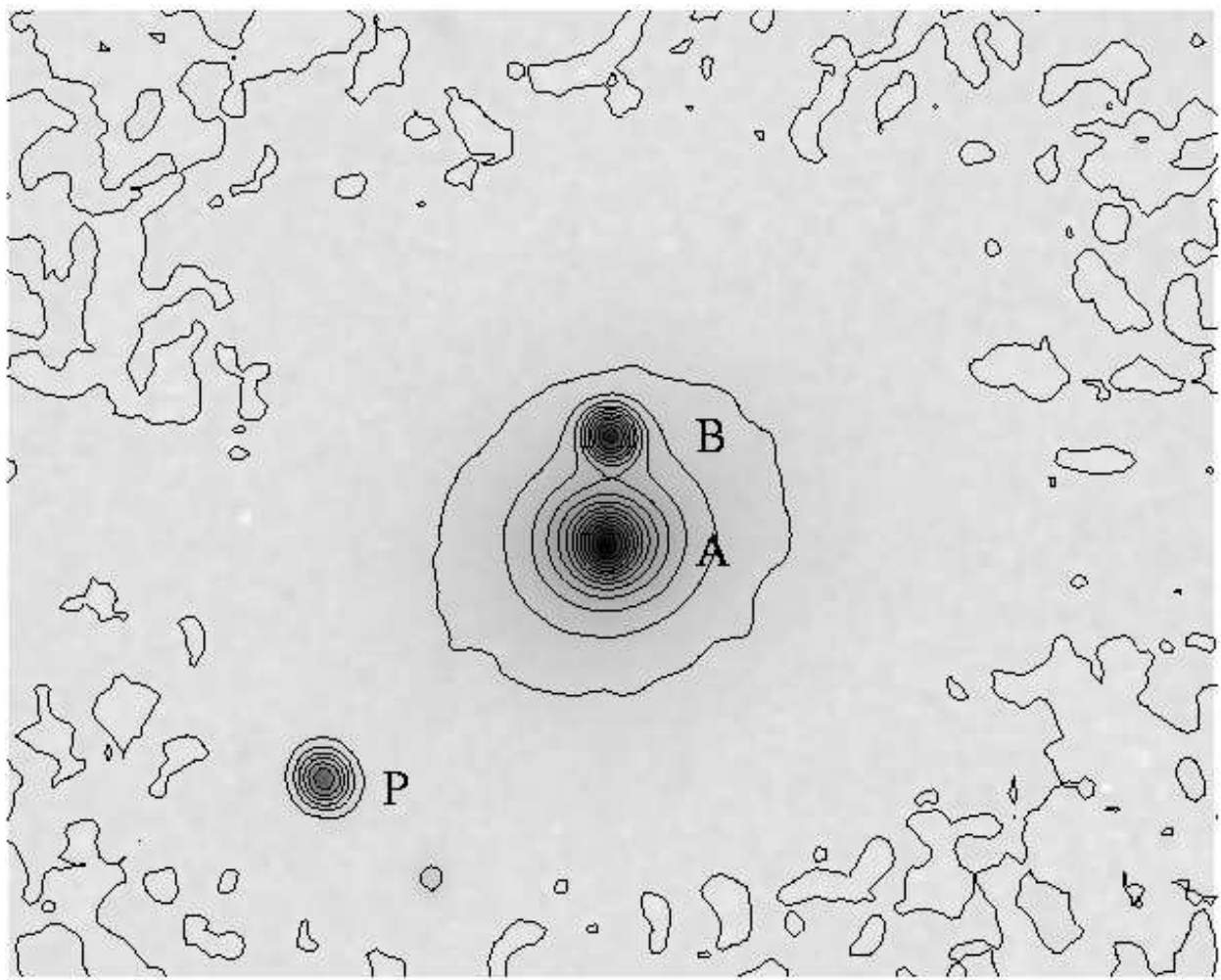,width=\textwidth}
	\caption{NIRC $K^{\prime}$-band image of SDSS~J1155+6346. Components A and B labelled indicating which component corresponds to which spectrum in Figure 1. Flux contours have been added to emphasize the extended profile of component A. Also indicated is P, the star used to model the half-width of the model point spread function. Up is 101$^{\circ}$ west of north.}
\end{figure}

\begin{figure}
        \center\epsfig{figure=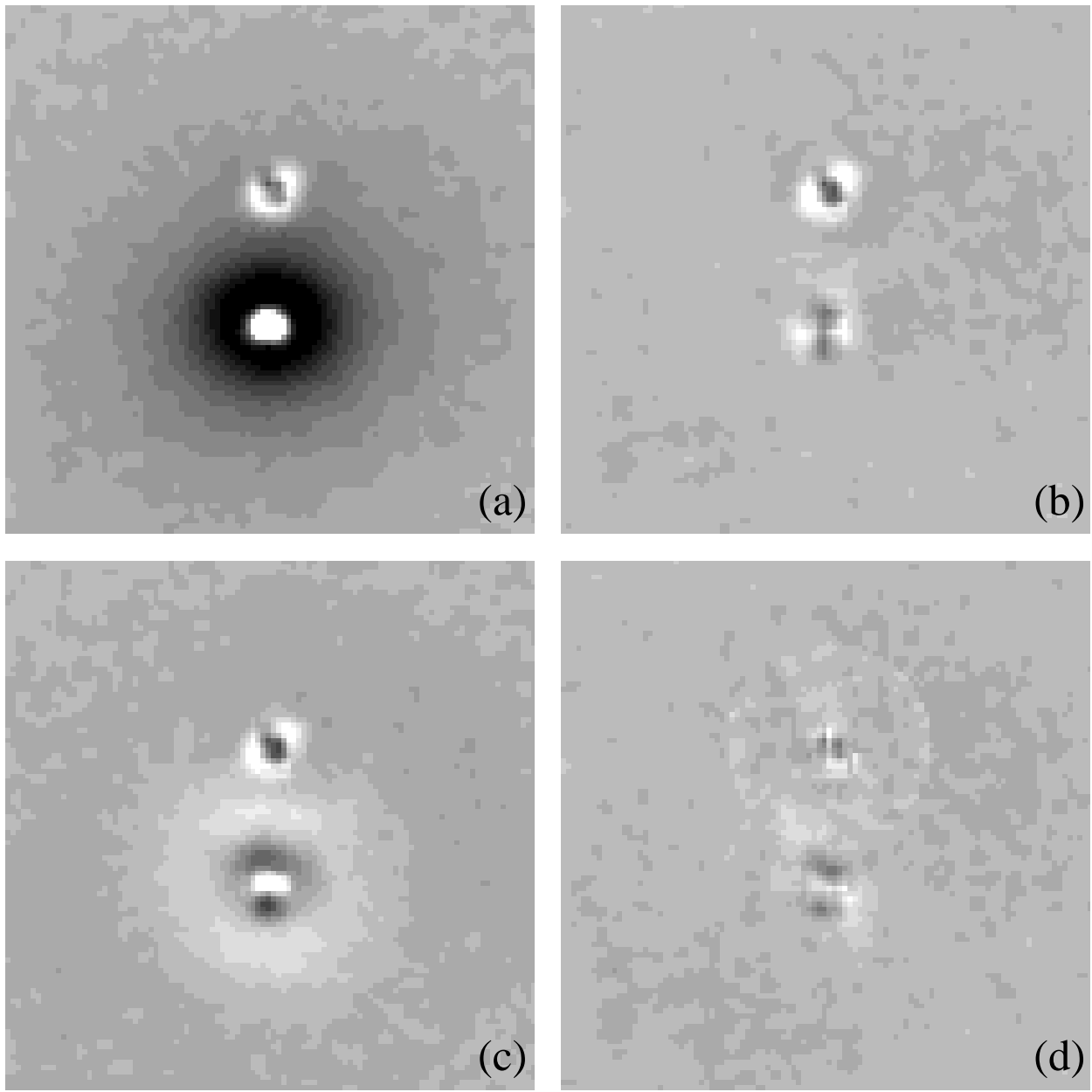,width=\textwidth,angle=-90}
        \caption{The remaining residuals after subtracting the best-fit model consisting of (a): two analytic PSFs. (b): two analytic PSFs plus a deVaucouleurs profile convolved with the PSF. (c): two analytic PSFs plus an exponential disc convolved with the PSF. (d): two empirical PSFs plus a deVaucouleurs profile convolved with the PSF. The extended ring of residuals shows where the empirical PSF has been truncated to mask out other sources, as explained in the text. The greyscale is the same in all four panels, but is different from the greyscale in Figure 4.} 
\end{figure}

\begin{figure}
	\centering\epsfig{figure=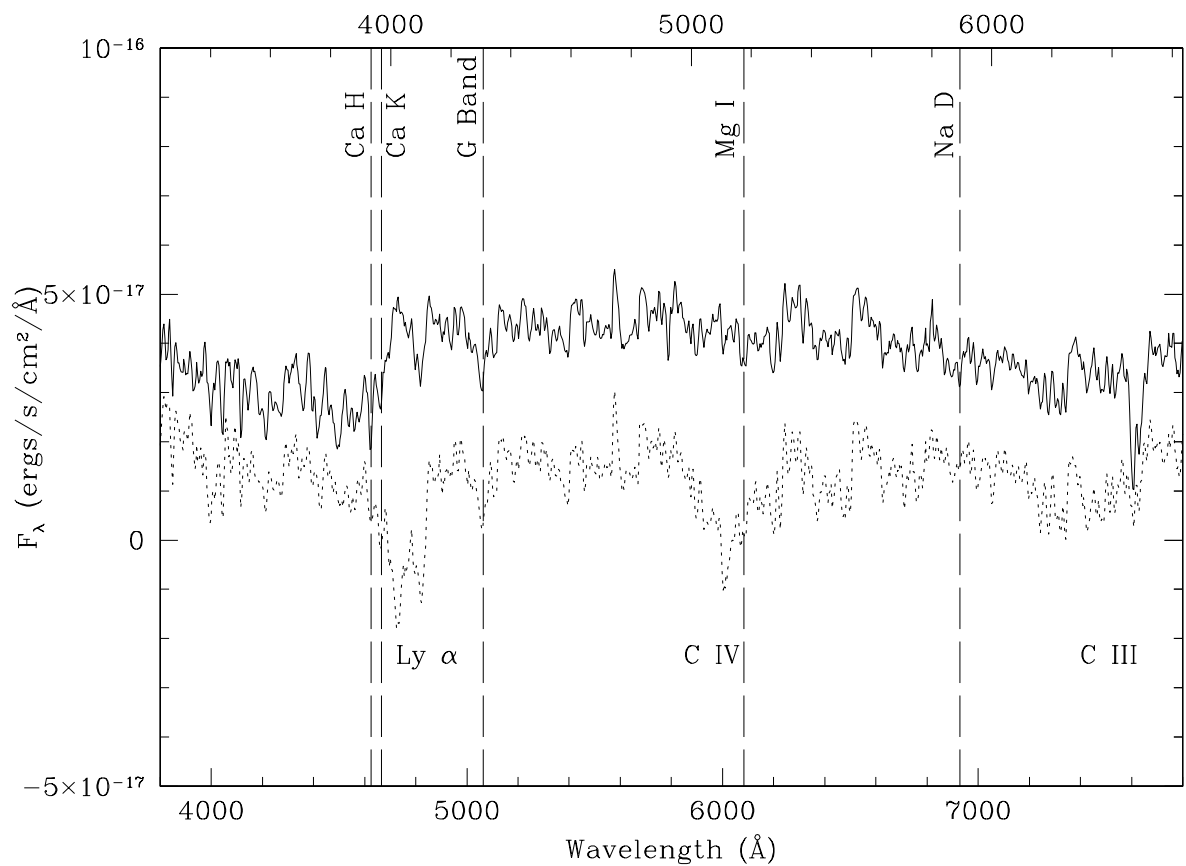,width=\textwidth}
	\caption{Difference of spectra for components A and B. The dotted curve is unscaled difference A$-$B. The solid curve is difference A$-$(B $\times$ ratio of A to B as determined by photometric modelling of NIRC image). The dashed vertical lines and vertical labels indicate prominent galactic absorption features at a redshift of $z=0.1756$, as described in Figure 2. The lower wavelength scale denotes the observed wavelengths and the upper scale denotes rest wavelengths corresponding to a redshift of 0.1756. The lower horizontal labels indicate the regions of quasar emission lines at a redshift of $z = 2.89$.}  

\end{figure}

\clearpage

\begin{deluxetable}{cccc}
\tablewidth{0pc}
\tablecaption{Photometric Model Parameters for SDSS~J1155+6346}
\tablehead{
\colhead{} & \colhead{Quasar A} & \colhead{Quasar B} & \colhead{Galaxy}
}
\startdata
\cutinhead{Two Analytic PSFs plus deVaucouleurs}
\textbf{Relative RA ($^{\prime\prime}$)} & 0 & -1.795 $\pm$ 0.007 & -0.14 $\pm$ 0.01 \\
\textbf{Relative Dec ($^{\prime\prime}$)} & 0 & -0.364 $\pm$ 0.004 & -0.062 $\pm$ 0.006 \\
\textbf{Relative Flux ($K^{\prime}$-band)}  & 1 & 2.0 $\pm$ 0.2 & 10.2 $\pm 0.6$\tablenotemark{a} \\
\textbf{$K$-Band Magnitude}\tablenotemark{b} & 17.4 $\pm$ 0.1 & 16.6 $\pm$ 0.1 & 14.9 $\pm$ 0.1\tablenotemark{a}\\
\textbf{$R_{\rm eff}$ ($^{\prime\prime}$)}  &  &  & 2.2 $\pm$ 0.1\\
\textbf{Axis Ratio ($b/a$)}  &  &  & 0.89 $\pm 0.02$\\
\cutinhead{Two Analytic PSFs plus Exponential Disc}
\textbf{Relative RA ($^{\prime\prime}$)} & 0 & -1.728 & -0.063 \\
\textbf{Relative Dec ($^{\prime\prime}$)} & 0 & -0.343 & -0.047 \\
\textbf{Relative Flux ($K^{\prime}$-band)}  & 1 & 0.9 & 1.3 \tablenotemark{a} \\
\cutinhead{Two Empirical PSFs plus deVaucouleurs}
\textbf{Relative RA ($^{\prime\prime}$)} & 0 & -1.792 & -0.188 \\
\textbf{Relative Dec ($^{\prime\prime}$)} & 0 & -0.377 & -0.220 \\
\textbf{Relative Flux ($K^{\prime}$-band)}  & 1 & 2.0 & 8.6 \tablenotemark{a} \\

\enddata
\tablenotetext{a}{Within measured effective radius}
\tablenotetext{b}{Calibrated to 2MASS photometry}
\end{deluxetable}
  
\end{document}